\documentclass{ctr_summer}

\usepackage{ctrfont}
\usepackage{natbib}
\usepackage{undertilde}
\usepackage[dvips]{graphicx}
\usepackage{psfrag}
\usepackage{epsfig}
\usepackage{amsmath,rotating}
\usepackage{dashrule}
\usepackage{subfig}
\usepackage[normalem]{ulem}

\usepackage{url}

\usepackage{color}

\def\xv{{\bf x}}

\def\xv{\mathbf{x}}

\def\U{\mathbf{U}}
\def\u{\mathbf{u}}

\def\bnabla{\boldsymbol{\nabla}}
\def\bcdot{\boldsymbol{\cdot}}

\title{Investigating nonlinearity in wall turbulence: regenerative versus parametric mechanisms}

\shorttitle{Regenerative or parametric mechanism}

\author{B. F. Farrell, E. Kim,\footnote[1]{School of Engineering and Applied Sciences, Harvard University}
H. J. Bae,\footnote[2]{Graduate Aerospace Laboratories, California Institute of Technology} M.-A. Nikolaidis \and P. J. Ioannou\footnote[3]{Department of Physics, National and Kapodistrian University of Athens, Greece}}

\shortauthor{Farrell et~al.}

\begin{document}

\pagenumbering{gobble}
%% Setting the first page number.  Please leave as it is, at "1".
\setcounter{page}{1}

\maketitle

Both linear growth processes associated with non-normality of the mean flow 
and nonlinear interaction transferring energy 
among fluctuations  contribute to
maintaining turbulence.  However, a detailed understanding of the
mechanism by which they cooperate in sustaining the turbulent state is lacking.  
In this report, we examine the role of fluctuation-fluctuation nonlinearity by varying 
the magnitude of the associated term in the dynamics of Couette flow turbulence  to 
determine how this nonlinear component helps maintain and determine
 the structure of the turbulent state, and particularly whether this mechanism is parametric 
or regenerative.  Having determined that the mechanism 
supporting the fluctuation field in Navier-Stokes turbulence is parametric, we   then study the 
mechanism by which the fluctuation component of turbulence is maintained by parametric growth 
in a time-dependent mean flow by examining the parametric growth mechanism in the 
frequency domain using analysis of  the time-dependent resolvent. 

\hrule

  \section{Distinguishing turbulence regimes: regenerative  versus parametric} 
   
Turbulence regimes can be distinguished by how  nonlinearity participates 
in maintaining the turbulent state. In this section we describe a study that distinguishes between the parametric 
and regenerative regimes  by performing simulations in which   
the strength of the fluctuation nonlinearity  is varied. 

Consider  a time-dependent incompressible  flow in a 
parallel channel, comprising the streamwise constant flow, $\U(y,z,t)\equiv(U,V,W)$, and the streamwise varying flow, \\
$\u'(x,y,z,t) \equiv (u',v',w')$. Here, $U$ and $u'$ are the velocity components in the streamwise direction $x$, $V$ and $v'$ are the velocity
components in  the cross-stream direction $y$,
and $W$, $w'$ are the velocity components in  the spanwise direction $z$,  that evolve according to the equations

%\begin{eqnarray}
\begin{subequations}
\label{eq:NS}
\begin{align}
\partial_t\U + {\U  \bcdot \bnabla  \U }   + \bnabla  P-   \Delta \U/ R  =& {- \langle\u ' \bcdot \bnabla  \u '\rangle_x},
\label{eq:NSm}\\
 \partial_t\u '+  { \U  \bcdot \bnabla  \u ' +
\u ' \bcdot \bnabla  \U } + \bnabla   p' -   \Delta  \u ' /R  =&  - \alpha  \left ( \u ' \bcdot \bnabla  \u ' - \langle \u ' \bcdot \bnabla  
\u' \rangle_x \right ),
\label{eq:NSp} \\
 \bnabla  \bcdot \U  = 0, \ \ \ \bnabla  \bcdot \u ' = 0,
 \label{eq:NSdiv0}
\end{align}
\end{subequations}
%\end{eqnarray}
with no-slip boundary conditions at the channel walls and periodic boundary conditions in $x$ and $z$.  
In these equations,  the streamwise 
mean component of a flow field is denoted by  $\langle \cdot \rangle_x$, and $\alpha$ 
has been included to allow variation in the influence of the
the non-linear interaction among the $\u'$ flow components.  The pressure
fields, $P(y,z,t)$ and $p'(x,y,z,t)$,  are determined from the incompressibility conditions,
and $R$ is a Reynolds number. In this report,  we study the  turbulence that develops in this system 
as a function of the parameter $\alpha$.
When $\alpha=1$, these equations are the Navier-Stokes equations for
the evolution of the total flow $\u = \U + \u'$, partitioned into  the two flow fields $\U$ and $\u'$ corresponding
to  the streamwise mean component  of $\u$ 
and  fluctuations of $\u$ from the streamwise mean.  Because of this identification, we refer to $\U$ as the mean
and  $\u'$ as the fluctuations for all values of $\alpha$. 

 Nonlinearity does not  act as an energy source in equations \eqref{eq:NS}  for any value of $\alpha$,
 and in  the absence of dissipation the energy of the total flow, $ \int_V d^3 \xv ~|\U + \u'|^2$,
is conserved for all values of $\alpha$.  Also, the fluctuation-fluctuation nonlinearity
does not contribute to the total energy of the fluctuation field, $\u'$, as   $\int_V d^3 \xv ~\u' \bcdot \left ( \u ' \bcdot \bnabla  \u ' - \langle \u ' \bcdot \bnabla  
\u' \rangle_x \right ) = 0$; therefore,  a non-vanishing fluctuation field 
is sustained  only by energy transferred between the mean $\U$ and the fluctuations $\u'$.
 This energy transfer arises from transient growth due to non-normality of the mean $\U$. 
The fluctuation-fluctuation nonlinearity, although it does not make a net contribution to the fluctuation energy,  can 
play a causal role in sustaining
the turbulence by redistributing energy among
the various fluctuation  structures, specifically by replenishing  the subset of fluctuations that 
participate in the non-normality induced transfer of
energy from the mean $\U$.
%{\color{red} We restrict the term regenerative mechanism to refer to mechanisms in which feedback from fluctuation-fluctuation nonlinearity 
%in the fluctuation equation is essential to maintaining the fluctuation variance.}
We restrict the term regenerative mechanism to refer only to 
 mechanisms in which this feedback from fluctuation-fluctuation nonlinearity 
in the fluctuation equation is essential to maintaining the fluctuation variance.
% this mechanism that relies on the fluctuation-fluctuation nonlinearity 
%to replenish the energy-extracting subspace of fluctuations by which the turbulence is  sustained 
% is referred to as the regenerative mechanism \citep{Jimenez-Moin-1991,Jimenez-2018}, and we restrict the term 
% regenerative mechanism to refer only to 
% these mechanisms in which feedback from fluctuation-fluctuation nonlinearity 
%in the fluctuation equation is essential to maintaining the fluctuation variance, in the streamwise mean-fluctuation decomposition of the Navier-Stokes equations}. 
It  has
recently been shown that this 
regenerative mechanism  can sustain a non-vanishing fluctuation field in the Navier-Stokes  equations
\eqref{eq:NSp} with $\alpha=1$
when the fluctuations interact with a specified time-independent, hydrodynamically 
stable, wall-bounded mean velocity profile  $\U(y,z)$, taken from a sufficiently non-normal streamwise mean flow
snapshot from a direct numerical simulation (DNS)   (Lozano-Dur\'an et al. 2021).

%\citep{Lozano-Duran-etal-2021}.

%\begin{figure}
%%\centering
%\begin{center}
%	    	\includegraphics[width = 0.75 \textwidth]{E_600_4p_newA.eps}
%      	\caption{ Time-averaged energies and 1-standard-deviation fluctuations (indicated with  bars)  of (a) the streak, (b) the roll, 
%	and (c) the fluctuations obtained from
%	equations \eqref{eq:NSm}-\eqref{eq:NSp} for different values of $\alpha$. (d) Time-averaged friction Reynolds number, $R_{\tau}$,  for the respective cases.  
%	The $R_{\tau}$ approaches its laminar value $\sqrt{R}$ (dashed line) as the parameter $\alpha$ is increased. 
%	At large enough values of $\alpha$, the flow becomes laminar. The vertical dashed lines indicate  Navier-Stokes turbulence at
%	 $\alpha=1$.  
%	 %The parameters of the simulation are given in Table \ref{table:geometry}.
%	  \label{Fig1}}
%	\end{center}
%\end{figure}

When  $\alpha=0$,
this regenerative mechanism is not available to sustain turbulence.  Nevertheless, 
not only is turbulence  sustained but  also the turbulence that develops, referred to as RNL turbulence, has
realistic structure, akin to the 
corresponding turbulence with $\alpha=1$,  
even at large Reynolds numbers (Thomas et al. 2014; Bretheim et al. 2015; Farrell et al. 2016, 2017a).
%\citep{Thomas-etal-2014,Bretheim-etal-2015,Farrell-etal-2016-VLSM,Farrell-etal-2016-PTRSA}.
This  RNL turbulence sustains its fluctuations by non-normal interaction with the time-dependent mean. This mechanism,  which
underlies the sensitive dependence on initial conditions associated with positive Lyapunov exponents,
and the universal exponential growth mechanism of perturbations  in  random matrix theory identified
by \cite{Oseledets-1968}, will be referred to as the parametric mechanism.
% \citep{Farrell-Ioannou-2012} 
%is referred to as the parametric mechanism \citep{Farrell-Ioannou-2012}.
Key  to the maintenance of the fluctuations  by the parametric mechanism is the time dependence of the mean flow,
which results in the subspace of the growing fluctuations
being continuously replenished in the absence of feedback regeneration by fluctuation-fluctuation nonlinearity.

\begin{figure}%[!tbp]
  \centering
  \subfloat[]{\includegraphics[width=0.56\textwidth]{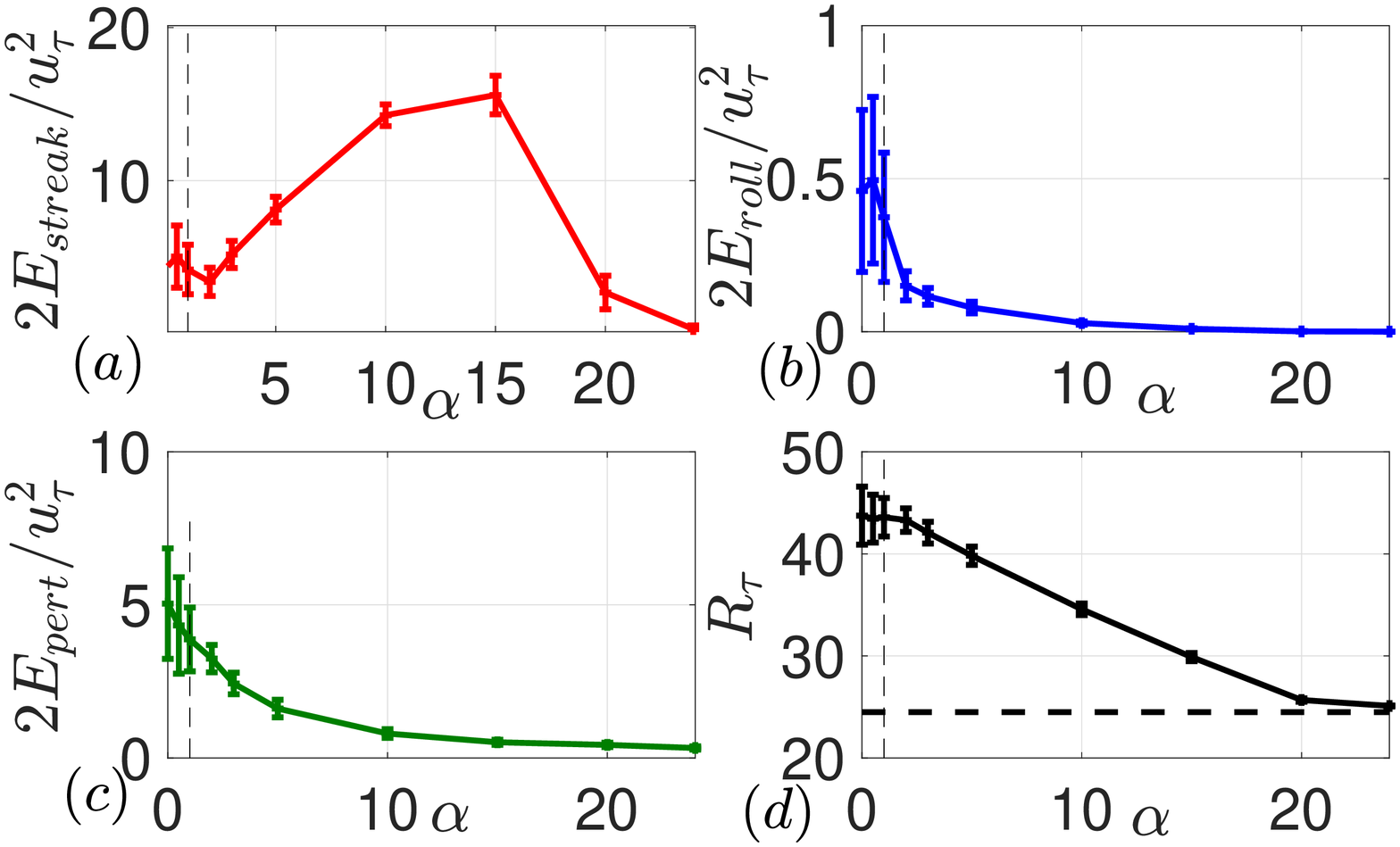}\label{fig:f1}}
  \hfill
  \subfloat[]{\includegraphics[width=0.4\textwidth]{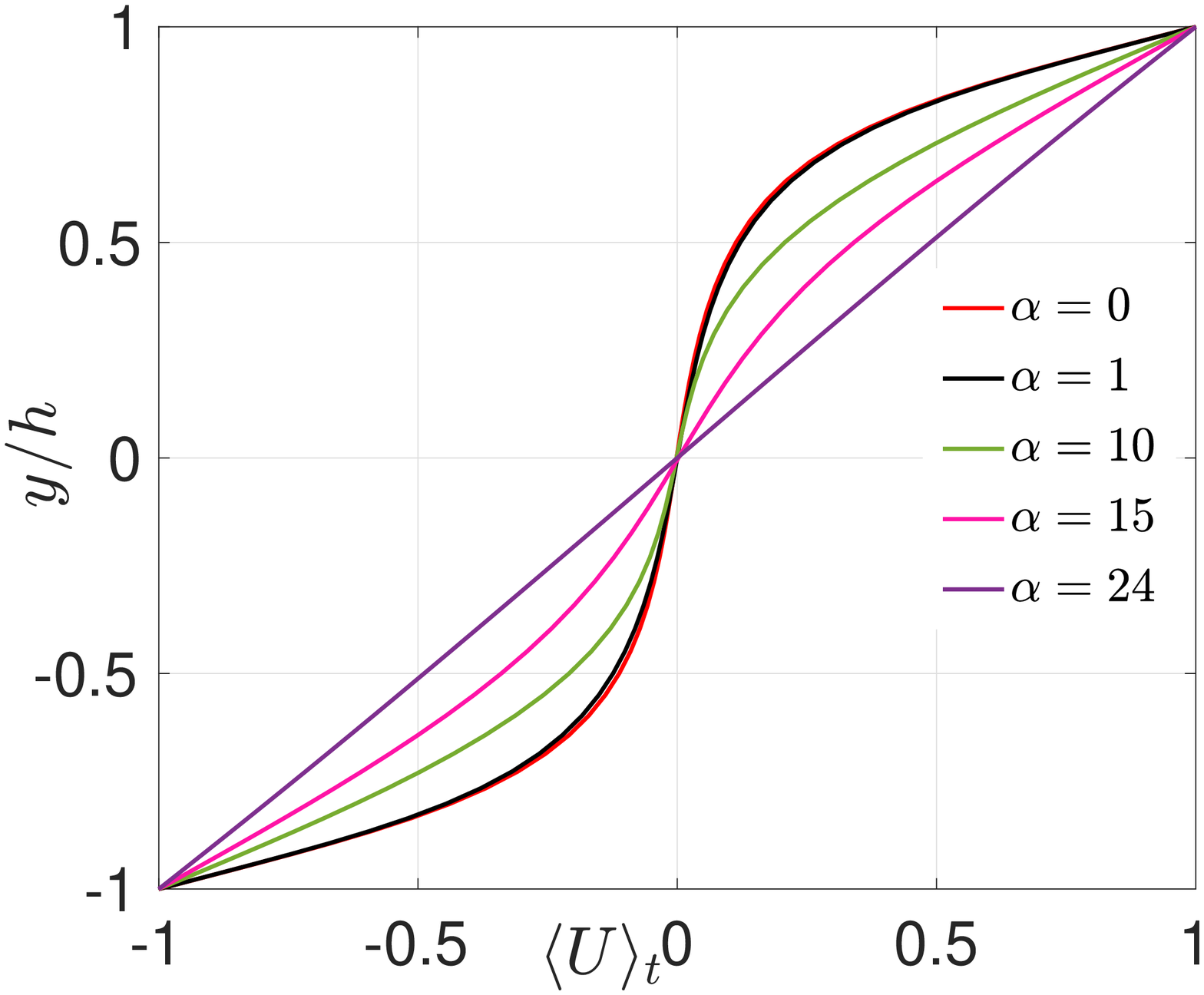}\label{fig:f2}}
  \caption{(I) Time-averaged energies and 1 standard deviation fluctuations (indicated with  bars)  of (a) the streak, (b) the roll, 
	and (c) the fluctuations obtained from
	equations \eqref{eq:NSm}-\eqref{eq:NSp} for different values of $\alpha$. (d) Time-averaged friction Reynolds number, $R_{\tau}$,  for the respective cases;  the dashed line indicates its laminar value  $\sqrt{R}$.
	%	The $R_{\tau}$ approaches its laminar value(dashed line) as the parameter $\alpha$ is increased. 
	%At large enough values of $\alpha$, the flow becomes laminar. 
	The vertical dashed lines indicate  Navier-Stokes turbulence at
	 $\alpha=1$. (II) Time-averaged streamwise mean flow for various values of $\alpha$. For $\alpha >20$, the flow is laminar.	 %The parameters of the simulation are given in Table \ref{table:geometry}.
	  \label{Fig1}  }
\end{figure}

%\begin{figure}%[!tbp]
%  \centering
%  \subfigure[I]{\includegraphics[width=0.56\textwidth]{E_600_4p_newA.eps}\label{fig:f1}}
%  \hfill
%  \subfigure[II]{\includegraphics[width=0.4\textwidth]{Up_a1p0_24p0_600.eps}\label{fig:f2}}
%  \caption{Time-averaged energies and 1 standard deviation fluctuations (indicated with  bars)  of (a) the streak, (b) the roll, 
%	and (c) the fluctuations obtained from
%	equations \eqref{eq:NSm}-\eqref{eq:NSp} for different values of $\alpha$. (d) Time-averaged friction Reynolds number, $R_{\tau}$,  for the respective cases.  
%	The $R_{\tau}$ approaches its laminar value $\sqrt{R}$ (dashed line) as the parameter $\alpha$ is increased. 
%	At large enough values of $\alpha$, the flow becomes laminar. The vertical dashed lines indicate  Navier-Stokes turbulence at
%	 $\alpha=1$.  Time-averaged streamwise mean flow for various values of $\alpha$. For $\alpha >20$, the flow is laminar.	 %The parameters of the simulation are given in Table \ref{table:geometry}.
%	  \label{Fig1}  
%  My flowers.}
%\end{figure}

%\begin{figure}%[!tbp]
%  \centering
%  \subfloat{\includegraphics[width=0.56\textwidth]{E_600_4p_newA.eps}\label{fig:f1}}
%  \hfill
%  \subfloat{\includegraphics[width=0.4\textwidth]{Up_a1p0_24p0_600.eps}\label{fig:f2}}
%  \caption{My flowers.}
%\end{figure}
%

%\begin{figure}
%%\centering
%\begin{center}
%	    	\includegraphics[width = 0.3 \textwidth]{Up_a1p0_24p0_600.eps}
%      	\caption{ Time-averaged streamwise mean flow for various values of $\alpha$. For $\alpha >20$, the flow is laminar.
%	 \label{FigU}}
%	\end{center}
%\end{figure}

Having identified at $\alpha=0$ a turbulent regime that is maintained exclusively  by the parametric mechanism,
we wish to examine the mechanism sustaining  turbulence as $\alpha$ increases to determine whether  turbulence 
can also be sustained   exclusively  by feedback regeneration for some value of  $\alpha$.
%We note that this would constitute a significant departure from the case of turbulence sustained with an
% imposed stable flow
%studied by \cite{Lozano-Duran-etal-2021}. 
Also, we wish  to determine whether the principal support of  Navier-Stokes turbulence at $\alpha=1$ is the parametric or 
the feedback regenerative mechanism.

We examine the turbulence that develops as a function of $\alpha$ in DNS  in the case of a Couette channel flow
at $R=600$
%The simulations are performed with an in-house developed DNS code  that solves equations \eqref{eq:NSm} and \eqref{eq:NSp}  in 
%wall-normal velocity/vorticity formulation \citep{Kim-etal-1987}. The code employs Chebyshev discretization on the wall-normal direction and a finite difference grid on the streamwise and spanwise directions which are treated pseudo-spectrally and are dealiased following the $2/3-$rule. Time-stepping is accomplished with a Crank-Nicolson/3rd order Runge-Kutta scheme for the viscous and advective terms respectively. Parameters of the simulations are summarized in Table \ref{table:geometry}. 
and consider the  dependence on $\alpha$ of  the friction Reynolds number,
$R_\tau$;
dependence on $\alpha$ of the energy of the streak component of the flow,
defined as $E_s(t)=\int_V d^3 \xv ~( U-\big \langle U \big \rangle_z)^2$, with $\big \langle U \big \rangle_z$ denoting
the spanwise averaged of the $U$ velocity; dependence on $\alpha$ of the energy of the  roll component of the flow,
$E_r(t)=\int_V d^3 \xv~ ( V^2 +W^2) $;
and also dependence on $\alpha$ of  the energy of the fluctuations, $E_p = \int_V d^3 \xv~ |\u'|^2 $.  In Navier-Stokes turbulence ($\alpha=1$) these
flow components underlie the self-sustaining cycle (SSP), which  is central to the maintenance of wall turbulence  
(Hamilton et al. 1995),  %\citep{Hamilton-etal-1995},  
and are  used   as diagnostics of the turbulent regime 
as
$\alpha$ varies.  

\begin{figure}
%\centering
\begin{center}
	    	\includegraphics[width = 0.55 \textwidth]{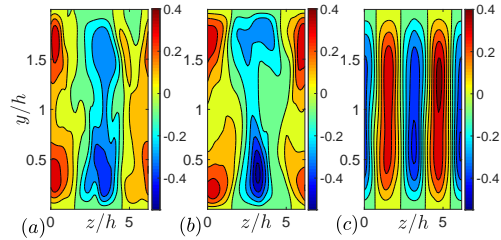}
      	\caption{ Snapshots of the streak. (a) in RNL turbulence for $\alpha=0$, (b) in Navier-Stokes  turbulence, $\alpha=1$, 
	and (c) in the almost stationary state  of turbulence at $\alpha=10$. ~~~~~~~~~~\label{fig:struct}} 
	\end{center}
\end{figure}

Three fluctuation turbulence regimes can be distinguished using these diagnostics, as shown in Figure \ref{Fig1}. The first regime extends
from
$\alpha=0$ to about $\alpha=2.5$ and in this regime the frictional Reynolds number together 
  with diagnostics of streak, roll, and fluctuation variance, are found to be typical of Navier-Stokes turbulence.
  Within this regime lies Navier-Stokes turbulence at $\alpha=1$, as well 
as RNL turbulence at  $\alpha=0$. 
Time-mean streamwise velocity profiles maintained in this Navier-Stokes turbulence 
regime are nearly indistinguishable, as are the  time-dependent  streamwise mean turbulent states, 
streak snapshots  of which are shown in 
Figure \ref{fig:struct}(a,b).  
%Particularly important are the essentially indistinguishable time-dependent  streak structures 
%that characterize both RNL turbulence at $\alpha=0$ and Navier-Stokes turbulence  
%at $\alpha=1$.  
We conclude that the 
parametric turbulence regime that has been identified to necessarily support 
RNL turbulence at $\alpha=0$ continues to provide 
the dynamical mechanism supporting turbulence for all values of $\alpha<2.5$, including 
Navier-Stokes turbulence at $\alpha=1$.  This turbulent regime can be identified with the time-dependent
SSP described in Hamilton et al. 1995.
As $\alpha$ is increased above $\alpha \approx 2.5$, the 
character of the turbulence changes, with a secular increase in the streak amplitude together with a decrease in the 
roll and perturbation amplitude (Figures \ref{Fig1}(Ia,b,c)).  This progression marks the transition
toward a
state of turbulence that is sustained by an almost constant time mean flow and,  therefore,  with fluctuations that
are sustained exclusively by the feedback regenerative mechanism in the absence of an SPP cycle. 
This feedback regenerative turbulence state 
corresponds to an analytic fixed point  of 
the  dynamics of the cumulants of the flow closed at second order (S3T) (Farrell \& Ioannou 2012; Farrell et al. 2017b).
%\citep{Farrell-Ioannou-2012, Farrell-Ioannou-2017-bifur}.
 Figure \ref{fig:struct}(c) shows an example of this fixed-point state at $\alpha=10$.
 This state is similar to that obtained by Lozano-Dur\'an et al. 2021, %\cite{Lozano-Duran-etal-2021}
with the important distinguishing characteristic that this state is self-maintaining.
%while that of \cite{Lozano-Duran-etal-2021}  required to be supported by defining the streak.  
We conclude that the parametric regime is required for the maintenance of 
RNL and Navier-Stokes turbulence but that a turbulent state 
is also maintained for  $\alpha>2.5$,  indicating that maintaining turbulence by  the feedback regenerative mechanism is 
 possible.  As $\alpha$ is increased further beyond values supporting 
the fixed-point regenerative turbulence, the turbulence eventually laminarizes, as indicated by 
the approach of the frictional Reynolds number and mean flow diagnostics to 
laminar values (Figures \ref{Fig1}(Id),(II)). We note that as $\alpha$ increases, the spatial spectrum of
the turbulence flattens, and it becomes impossible to prevent  backscatter to the larger scales from
the smallest scales retained in the DNS. For this reason, as shown in Figure \ref{Fig1}(Ic),  there remains a small 
 level of fluctuations at $\alpha>15$ in what we believe should 
 have been a strictly laminar state. 
 Corroboration is provided by the fact that S3T
at these values of $\alpha$,  with a stochastic parameterization of the fluctuation-fluctuation nonlinearity 
in which backscatter is not supported, do obtain the laminar state.

We conclude that the regime of turbulence that includes RNL and Navier-Stokes turbulence is 
maintained by the parametric mechanism of energy transfer from the mean to the fluctuation field
 that is isolated using RNL 
analysis.  This parametric turbulence regime
does not depend on  fluctuation-fluctuation nonlinearity for its support.  
%However,  
%a fluctuation-fluctuation nonlinearity-based mechanism has been shown to support a fixed-point turbulent state at $\alpha=10$. 
With the fluctuation-fluctuation nonlinearity  shown to be unnecessary for the 
maintenance of the fluctuation component of turbulence, we turn our attention to the remaining 
nonlinearity, which is the parametric transfer of energy from 
the time-dependent mean flow to the fluctuation field.  This transfer 
is associated with the Lyapunov vectors of the 
time-dependent mean flow.  
%Moreover, whether our interest is in the 
%maintenance of turbulent  fluctuations or in the influence of 
%fluctuations arising from nonlinearity or other sources on the turbulence,
%gaining a better understanding requires the study of perturbation growth in 
%time-dependent flows, which is the parametric growth process.

%For example in \cite{Lozano-Duran-etal-2021}
%the turbulence  sustained by the purely  regeneration mechanism with an imposed  mean flow
%so that the streak and roll energy are constant and the energy of the fluctuations varies with mean value  
%$E_p/u_\tau^2 \approx 2$.

\section{Resolvent analysis of time-dependent systems}

We continue our study of the role of nonlinearity in turbulence by studying 
the mechanism by which fluctuations are amplified by non-normal interaction 
with a time-dependent mean flow.   Our analysis method
is performed in the frequency domain and is based on  extension of 
resolvent analysis to time-dependent mean flows.   The formalism presented builds on  
that presented by   Wereley(1991), Fardad et al. (2008), Padovan et al. (2020), Rigas et al. (2021), and Franceschini et al. (2022).
%\cite{Wereley-1991-phd,Jovanovic-eta-2008,Padovan-etal-2020,Rigas-etal-2021,Franceschini-etal-2022}. 
We begin by recalling that turbulence in shear flow is sustained by transfer of energy from the large-spatial-scale externally 
forced flow field  to the small-scale fluctuation field and that this  transfer is mediated by the non-normality 
of the large-scale forced flow.  Non-normal interaction between large- and small-scale components of the turbulent 
flow field can be analyzed in either  the time or frequency domain, and both of these approaches 
are contained within generalized stability theory (GST) 
\citep{Farrell-Ioannou-1996a, Farrell-Ioannou-1996b,Farrell-Ioannou-1999,Schmid-Henningson-2001}. However, 
while transient growth analysis for time-dependent flows is well developed in the time 
domain, analysis of the GST of time-dependent flows in the frequency domain is less developed.  
This is  important because the large-scale streamwise mean flows involved in the dynamics of shear 
turbulence are highly time-dependent.
%Resolvent analysis in turbulent studies is predicated on the frequency response of perturbations on a time-independent mean flow.
Moreover, as we have seen, the fluctuation component of the turbulence
is maintained by parametric growth,
which motivates a closer 
analysis of the mechanism by which parametric growth takes place.

The  
dynamics of perturbations, $x$,  linearized about  a time-dependent mean state can be written, using a partition
into the time-mean, $A$, and perturbation, $B$, operators,  as 
\begin{equation}
\dot x = (A + B(t))x +f(t),
\label{eq:1}
\end{equation}
where $f(t)$ is
a  time-dependent excitation.  Assuming asymptotic stability of the system, by Fourier transforming  Eq. \eqref{eq:1}, we obtain
\begin{equation}
 -i \omega \hat{x}(\omega) = A  \hat{x}(\omega) +  \frac{1}{\sqrt{2\pi}} \int_{-\infty}^\infty d\omega'~ \widehat{B}(\omega-\omega') \hat{x}(\omega') +\hat{f}(\omega),
 \label{eq:recur0}
\end{equation}
where $\hat{x}(\omega)$ is the Fourier transform of the state $x(t)$ and $\widehat{B}(\omega)$ the transform
of the operator $B(t)$. 
Because of the convolution in Eq. \eqref{eq:recur0}  the temporal evolution of individual
free modes of this time-dependent system will not be characterized solely by 
the eigenvalues of $A$, and the 
asymptotic response of the system to an excitation 
at a single frequency will not be restricted to the frequency of the forcing.
Analysis of the dynamics in frequency space governed by Eq. \eqref{eq:recur0} proceeds by defining 
%PROCEEDS BY DEFINING  and the   study of the response requires
%that we broaden the  response to include  all frequencies.
%For that we define 
an operator, $H$, acting on all the Fourier components of $\hat{x}$, as 
\begin{equation}
\left . H \hat{x}\right |_\omega \equiv  i \omega \hat{x}(\omega) + A  \hat{x}(\omega) +  \frac{1}{\sqrt{2\pi}} \int_{-\infty}^\infty d\omega'~ B(\omega-\omega') \hat{x}(\omega').
\label{eq:Hill}
\end{equation}
We refer to this operator as the Hill operator. The frequency response 
of the time-dependent dynamics (Eq. \eqref{eq:1}) to excitation $f(t)$  
is given in terms of the resolvent of the Hill operator, 
$R=-H^{-1}$, which  determines the response, $\hat{x} =R \hat{f}$.  The eigenfunctions of
$R$, which are also the eigenfunctions of $H$, are the resonant responses of this time-dependent system.
%The convolution in \eqref{eq:recur0} accounts for the
%interactions  of the Hill {\color{red} eigenmodes} with the component of the time-dependent dynamics.
%\sout{which are not transparent in the time-domain formulation of the dynamics, and because the convolution has usually
%narrow bandwidth  approximations based on the Hill matrix formulation have advantageous convergence 
%properties compared to those based on the formulation of the dynamics in the time domain.}

\subsection{Physical meaning and properties of the spectrum of the Hill operator}

Consider the linear dynamics
\begin{equation}
\dot x = (A+B(t)) x.
\label{eq:1a}
\end{equation}
The eigenfunctions of $H$ can be identified with the time development of
the covariant Lyapunov vectors (CLVs)  of this system and 
the real part of the eigenvalues of $H$ with the Lyapunov exponents
of the corresponding CLVs.  
As an example consider   $B(t)=B\cos t$. In this periodic case,
we know from Bloch's theorem that the eigenfunctions of the dynamics, i.e. the CLVs,  are of the form
$x_\lambda (t) = e^{\lambda t} \chi_\lambda (t)$, where $\chi_\lambda$ is a periodic function with period $2 \pi$, and that
the real part of
the exponent $\lambda$ is the Lyapunov exponent of this CLV and the imaginary part of $\lambda$ is the mean rate of phase advance
of the periodic function $\chi_\lambda$.  The connection between the eigenfunctions of $H$ and the CLVs of Eq.  \eqref{eq:1a}
arises because the periodic function component of the CLV, $\chi_\lambda$,  is the Fourier transform of the eigenfunction
of $H$ with eigenvalue $\lambda$.  Indeed, if we write the periodic function $\chi_\lambda(t)$  as
a Fourier series, $\chi_\lambda(t)= \sum_{n=-\infty}^\infty \hat{x}_n e^{-i n t}$,
 the eigenfunction becomes $x_\lambda (t) = e^{\lambda t} \sum_{n=-\infty}^\infty \hat{x}_n e^{-i n t}$,
 and upon introducing  this functional form in Eq. \eqref{eq:1a},  we find, by matching the corresponding Fourier components
 that arise in the expansion of Eq. \eqref{eq:1a},  that the Fourier components $\hat{x}_n$ must satisfy for all $n$ the 
 recurrence relations
  \begin{equation}
i n \hat{x}_n + A \hat{x}_n + \frac{B}{2} \left ( \hat{x}_{n+1} + \hat{x}_{n-1} \right ) = \lambda \hat{x}_n.
\label{eq:floquet1} 
\end{equation}
The left side of Eq. \eqref{eq:floquet1} is the $n$-th Fourier component of $H \hat{x}$, with $H$ the Hill 
operator discretized on the integer-valued frequency lattice and Eq. \eqref{eq:floquet1}   is the statement that $\hat{x}$ is the eigenfunction of $H$ with eigenvalue
$\lambda$, which we have shown to be equivalent to the statement that the  CLV 
$x_\lambda (t) = e^{\lambda t} \chi_\lambda (t)$ satisfies Eq. \eqref{eq:1a}.  Note 
also from Eq. \eqref{eq:floquet1} that if $\hat{x}_n$ are the components of an eigenfunction
 of $H$ with eigenvalue $\lambda$, then for any integer $m$,  $\hat{y}_n=\hat{x}_{n-m}$,
 will also  be an  eigenfunction of $H$ with eigenvalue  $\lambda+i m$.
 However,  all these  shifted  eigenfunctions of $H$ correspond to the same  CLV as 
 $x_\lambda(t) = e^{\lambda t} \sum_{n=-\infty}^\infty \hat{x}_n e^{-i n t}=
  e^{\lambda t} \sum_{n=-\infty}^\infty \hat{x}_{n-m} e^{-i (n-m) t}=
  e^{(\lambda +im) t} \sum_{n=-\infty}^\infty \hat{y}_{n} e^{-i n t}$. The shifted eigenfunctions of  $H$ 
  arise because the periodic function $\chi(t)$, which is associated with the eigenfunction of $H$,   can be multiplied by 
an arbitrary exponential, $e^{imt}$, with the underlying period. All these shifted eigenfunctions  are required in order
to produce completeness of the representation in the frequency domain.

	 The above analysis carries over to the general case when $B(t)$ is not periodic. The CLVs of Eq. \eqref{eq:1a} are 
 then written as $x_\lambda (t) = e^{\lambda t} \int_{-\infty}^\infty d \omega ~\hat{x}(\omega) ~e^{- i \omega t}$,
 and upon introduction of this expression in Eq. \eqref{eq:1a},  we  find that the components of 
 $\hat{x}$  must satisfy 
 \begin{equation}
\left . H \hat{x}\right |_\omega \equiv  i \omega \hat{x}(\omega) + A  \hat{x}(\omega) +  \frac{1}{\sqrt{2\pi}} \int_{-\infty}^\infty d\omega'~ B(\omega-\omega') \hat{x}(\omega')  = \lambda  \hat{x}(\omega),
\label{eq:Hilleig}
\end{equation} 
 in other words  that $\hat{x}$  should be an eigenfunction of $H$ with eigenvalue $\lambda$. Also, as in the periodic case,
 if $\hat{x}_\lambda$ is an eigenfunction of $H$  with eigenvalue   $\lambda$
  so is $\hat{x}_{\lambda,\alpha}(\omega)=\hat{x}_\lambda(\omega-\alpha)$ with eigenvalue $\lambda + i \alpha$, obtained by shifting $\hat{x}_\lambda$ by $\alpha$,  and  all these eigenfunctions produce the time evolution of 
 the same CLV, $x_\lambda(t)$, 
in the time domain. Note that  Floquet analysis
determines the CLV only at a chosen time, while the Hill matrix approach obtains the full time 
evolution of the CLVs. 
\subsection{Variance maintained at steady state by  stochastic excitation 
and optimal response  to harmonic excitation of the time-dependent dynamics }

%  \begin{figure}
%%\centering
%\begin{center}
%	    	\includegraphics[width = 0.3 \textwidth]{VELt_e0p16.eps}
%      	\caption{Snapshots of the time-dependent mean velocity profile for $\varepsilon=0.16$ with
%	the resolution of  $N_y=21$ levels used in the calculations. Despite the coarseness in the resolution the salient features in the response are captured accurately.   \label{fig:velt}}
%	\end{center}
%	\end{figure}

 We want to determine the modification 
 to the steady-state response to stochastic excitation resulting from time dependence of the system dynamics.
 In the energy metric, this response can be conveniently obtained using 
 the ${\cal L}^2$ norm $|| \hat{x} ||^2 = \int_{-\infty}^\infty d \omega ~|\hat{x}(\omega)|^2$
 by first transforming $\hat{x}$   into energy coordinates so that this ${\cal L}^2$ norm corresponds to energy. 
  Assuming that the time-dependent operator, $A+B(t)$, is stable, the ensemble  mean energy density at steady 
 state is % assuming the frequencies have discretized and assume integer values of $\delta \omega$,  is 
 $\delta \omega ~{\rm trace}\left ( \langle \hat{x} \hat{x}^\dagger \rangle \right )$, which is 
$\delta\omega~ {\rm trace}\left ( R \langle \hat{f}\hat{f}^\dagger \rangle R^\dagger \right )=\delta\omega/(2\pi)~{\rm trace}\left ( R R^\dagger \right )\equiv \delta\omega/(2\pi))||R||_{F}^2$,  where $||R||_{F}$ is the Frobenius norm of the Hill resolvent
and $\delta \omega$ is the increment in frequency of the Riemann sum approximation of the integral in the ${\cal L}^2$ norm. 
We have   
assumed  whiteness of the stochastic excitation, that is, that
the spatial components of the forcing satisfy $\langle \hat{f}_i(\omega) \hat{f}_j^*(\omega') \rangle =  \delta_{ij} \delta_{\omega,\omega'}/(2 \pi)$. 
When the dynamics is discretized on a finite grid,  $||R||_F$ is  the square root of the sum of the squares of the singular values of $R$.
In the time-independent case it is useful to compare the variance maintained by the system to the variance that would  be maintained
if the system were normal and therefore the modes of the system were orthogonal and as a consequence contribute
to the variance without intermodal interactions.
The variance of this equivalent normal system per unit forcing is given by the classical resonant response to forcing, $||R_{eq}||_F=\sum_i 1/(2 \lambda_i)$, where $\lambda_i$ are the decay rates of the modes.  By necessity $||R_{eq}||_F\le ||R||_F$  \citep{Ioannou-1995}, and the ratio $||R||_F/||R_{eq}||_F$ can serve as a measure of the non-normality of the dynamics. In  time-dependent dynamics we can similarly
define  the variance of the effectively normal system, $||R_{eff}||_F=\sum_i 1/(2 \lambda_i)$, where now the $\lambda_i$ are the Lyapunov exponents 
of the time-dependent dynamics. 
We could also consider  $||R(\varepsilon)||_F/||R_{eff}(\varepsilon)||_F$ as a measure of the mean non-normality of the CLVs.
%Clearly, $||R_{eff}(0)||_F=||R_{eq}||_F$.% when $\varepsilon=0$. }
%The  contribution of  non-normality of  $H$ will increase the maintained variance over that of the equivalent normal response associated with $H$,   . 

In addition to the variance, it is  of interest to obtain the optimal response that could result from a 
deterministic harmonic excitation at frequency $\omega_e$.  This is given by  the ${\cal L}^2$ norm
of $R P_{\omega_e}$, where $P_{\omega_e}$ is the operator that projects $\hat{f}$ to its value at the excitation  frequency $\omega_e$. When the dynamics is discretized on a finite grid, it is given by the largest singular value  of 
$R P_{\omega_e}$.  In the time independent case we can similarly define the equivalent normal response 
to be the response that obtains when the modes of $A$
are assumed to be orthogonal.

\begin{figure}
%\centering
\begin{center}
	    	\includegraphics[width = 0.6 \textwidth]{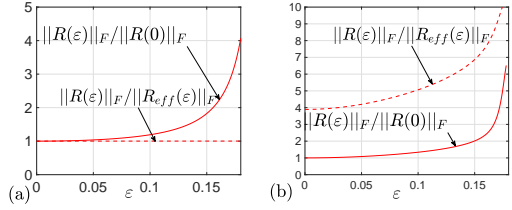}
      	\caption{Shown as a function of the amplitude of the time dependence
$\varepsilon$ are the root-mean-square variance, $||R||_F$, 
	maintained under white stochastic excitation (solid lines), and the 
	ratio of the variance maintained by the time-dependent system to the variance, $||R_{eff}||_F$, maintained by the 
	effectively  normal  dynamics (dashed lines). %which maintains variance proportional to
	%$\sum_i 1/(2\lambda_i)$, where $\lambda_i$ are the  Lyapunov exponents of the time-dependent dynamics.
	(a) Results for  the Mathieu equation with  $\omega_f=2 \omega_0$,  $\omega_0=1$, and $\gamma=0.05$. 
	The oscillator becomes unstable with these parameters at $\varepsilon_c=0.2$.
	(b) Results for the time-modulated Poiseuille flow at $R=800$ excited with perturbations with wavenumbers
	 $k=1.143$, and $m=1.67$. 
	 The time-dependent Poiseuille flow 
	becomes unstable at $\varepsilon_c=0.178$.
	%The solid curve shows the increase in variance due to the time dependence of the operator while the 
	%dashed curve shows the contribution to this increase arising from the increase in non-normality 
	%arising from the time dependence.  	
	%In the Mathieu equation at $\omega_f=2 \omega_0$
	%the response to stochastic forcing is as if the covariant Lyapunov vectors were normal modes of the time dependent system 
	%(this is not true at other $\omega_f$).
	\label{fig:Fig2}}
	\end{center}
	\end{figure}

\begin{figure}
%\centering
\begin{center}
	    	\includegraphics[width = 0.5 \textwidth]{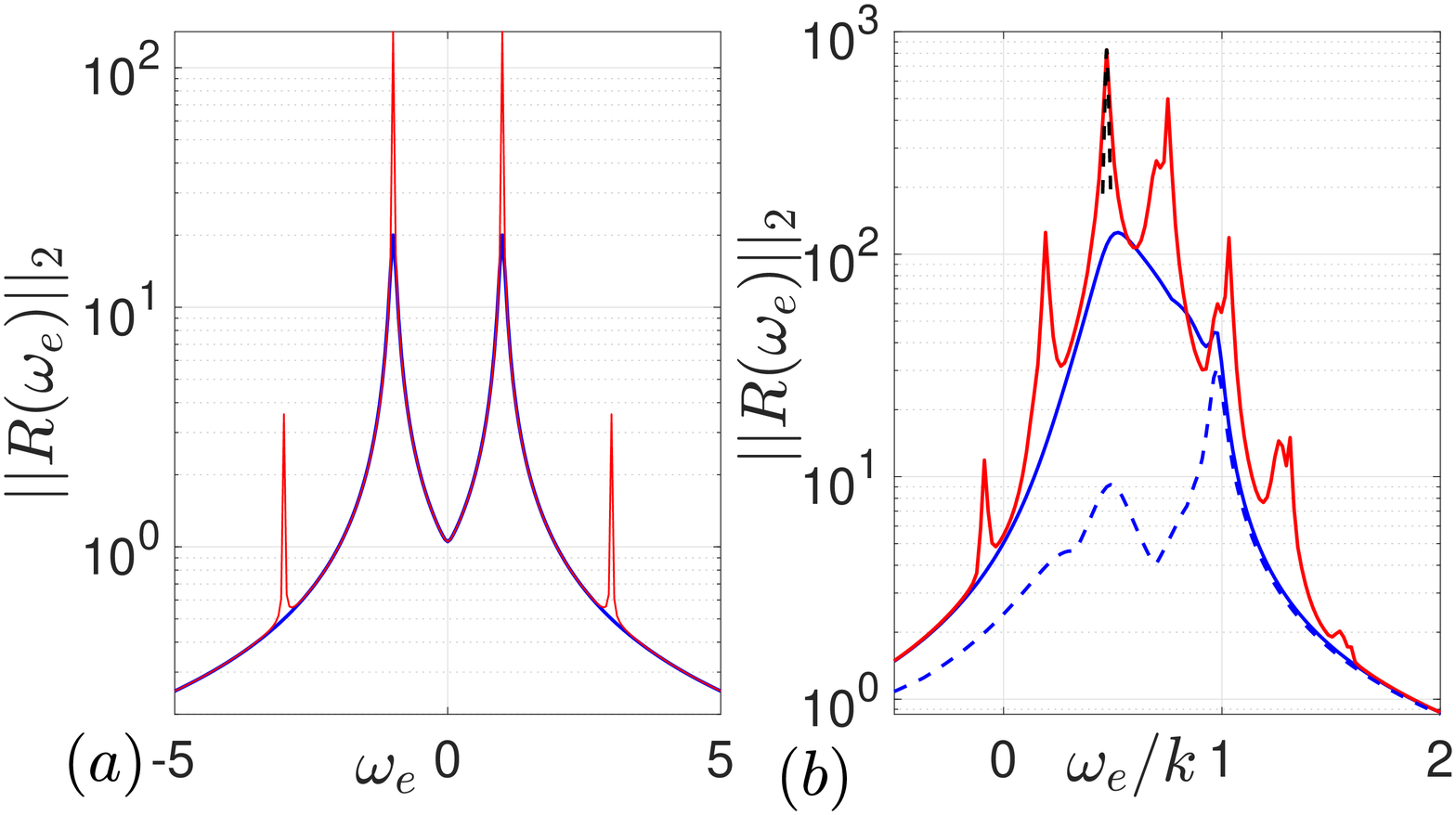}
      	\caption{The optimal response $||R(\omega_e)||_2$ 
	%of two time-dependent systems to harmonic forcing 
	 as a function of  $\omega_e$.  
		(a) Results for the Mathieu equation with  $\omega_f=2 \omega_0$,  $\omega_0=1$, and $\gamma=0.05$. The blue line
		is the response for $\varepsilon=0$, and  the red line is the response for $\varepsilon=0.18$. 
	(b) Results for the time-modulated Poiseuille flow. The solid blue line represents the response %of this non-normal system
	 without time dependence ($\varepsilon=0$), the dashed blue line represents
	 the equivalent normal response, $||R_{eq}||_2$, of the time-independent $A$ operator. 
	 The red line represents the response with time dependence  ($\varepsilon=0.16$ and $\omega_f=0.3$). 
	%Green line: $\varepsilon=0.16$ and $\omega_f=0.1$.
	The peaks are separated by  $\omega_f$ and their width is determined by the Lorentzian (dashed black curve) 
	corresponding to the Lyapunov exponent.
%	the Lyapunov exponent as 
%	The dashed black curve at the frequency of maximum response   at $\omega_f=0.3$ is the Lorentzian  with decay rate equal to the Lyapunov exponent. %Agreement locally    of the Lorentzian and the response curve verifies that the response 
	%is determined by the least stable Lyapunov exponent.   
\label{fig:Fig3}}
	\end{center}
\end{figure}

 Consider the damped Mathieu equation, which is the prototypical example of parametric instability, with
\begin{eqnarray}
A = 
\left(
\begin{array}{ccc}
 0  & 1  \\
-\omega_0^2  & - 2 \gamma  
\end{array}
\right),~~~B(t)=
 \varepsilon \cos(\omega_f t) B,~~B=
\left(
\begin{array}{ccc}
 0  & 0  \\
1  & 0   
\end{array}
\right),
\label{eq:mat}
\end{eqnarray}
where $\gamma=0.05$, $\omega_0=1$, and $\omega_f=2$. These  parameters correspond to forcing at twice the resonant frequency,
which can be shown to  lead to maximal amplification.
The associated Hill operator becomes  
\begin{equation}
\left . H \hat{x} \right |_\omega =  i \omega \hat{x}(\omega)  + A  \hat{x}(\omega) +  
\frac{\varepsilon}{2} B (  \hat{x} (\omega-\omega_f) +  \hat{x}(\omega+\omega_f) ),
\label{eq:Hillcos}
\end{equation}
which in the calculations will be discretized on a finite  lattice of frequencies. 
Figure \ref{fig:Fig2}(a) shows the variance maintained under stochastic excitation as a function of $\varepsilon$.
As expected, the maintained variance increases as $\varepsilon$ approaches the critical  $\varepsilon_c$ 
at which the dynamics become unstable ($\varepsilon_c=0.2$ for these parameters).  
Also shown is the effective non-normality of the dynamics measured by  $||R(\varepsilon)||_F/||R_{eff}(\varepsilon)||$.
%the ratio of the maintained variance to the
%variance, $||R_{eff}||=\sum_i 1/(2\lambda_i)$, where $\lambda_i$ are the Lyapunov exponents, 
%maintained in the effectively normal dynamics, that is the variance that 
%would be maintained  if the CLVs were considered as  normal coordinates. 
At these parameter values,
the maintained variance increases at the rate 
predicted by the inverse sum of
the Lyapunov exponents of the two CLVs,
as if the CLVs of the Mathieu equation were  normal coordinates. This is reflected  in the minimal increase of the effective non-normality of the time-dependent dynamics 
as a function of $\varepsilon$,  shown by 
the dashed line in  Figure \ref{fig:Fig2}(a).
Note also that this figure verifies that the Mathieu equation is normal at $\varepsilon=0$.
The variance  maintained by excitation of  the parametric Mathieu oscillator is also close to the variance
of the equivalent normal system at other parameter values. 
Figure  \ref{fig:Fig3}(a) shows that the
optimal response to harmonic forcing
with $\varepsilon=0.18$ and $\varepsilon=0$
is  maximized in both cases for forcing at the resonant frequencies $\omega=\pm\omega_0$.
Also, given that $\omega_f=2 \omega_0$, there is strong response 
at the sideband frequencies  $\omega=\pm 3\omega_0$. 
%The response at frequencies higher than $\omega_0 +  k \omega_f$, for $k$ an integer is  not discernible. 
We conclude that,
in the case of the asymptotically stable  Mathieu equation, time dependence produces 
an  enhanced response primarily due to a  decrease of the decay rate of the Lyapunov exponents of the system.
%while the system is
%normal in the mean and in  the mean the CLVs of the Mathieu equation were normal coordinates.

In contrast to the Mathieu oscillator, which is characterized by 
normal mean dynamics,  the streamwise mean of a shear flow is typically
highly non-normal,  even when the streamwise mean flow is time-independent. 
For example, consider the resolvent analysis for
plane wave perturbations 
in a plane parallel channel in $y\in [-1,1]$ with  streamwise mean flow $U(y,t)=U_0(y)+\varepsilon \cos(\omega_f t) U_1(y)$, where 
$U_0(y)=1-y^2$ and $U_1(y)=\sin(2 \pi y)$, and with forcing frequency
$\omega_f=0.3$, streamwise wavenumber $k=1.143$, and spanwise wavenumber 
$m=1.67$ at Reynolds number $R=800$.   The excitation frequencies $\omega_f$ have been chosen 
to lie in the interval for which  time dependence leads to an amplified response,  which is for  frequencies in the range 
$0.08<\omega_f < 0.35$. 
%Snapshots of the fluctuating mean
%velocity profile for $\varepsilon = 0.16$ are shown in Fig. \ref{fig:velt}. 
The perturbations evolve according to the
Orr-Sommerfeld-Squire (OSS) equations,  where $A$ and $B$ are, respectively, 
the time-independent and time-dependent components of the OSS operator. The complex $A$  and $B$ 
operators  are discretized 
and   expressed in energy coordinates as in  \cite{Butler-Farrell-1992}.  
%The results we present are  with a discretization with
%$N_y=21$ levels in $y$.  Despite the coarseness in the resolution  it has been verified that 
%the salient features in the response are captured accurately.   
 Figure \ref{fig:Fig2}(b) shows the maintained variance  and  the effective mean  non-normality   of the time-dependent dynamics
 as a function of $\varepsilon$ 
for the case of $\omega_f=0.3$.  Unlike in the case of the parametric oscillator,
both the variance and the effective mean non-normality increase with the amplitude of the time-dependent component of the dynamics.  
Figure  \ref{fig:Fig3}(b) shows the optimal response as a function of excitation frequency. 
The response is asymmetric with respect to the zero frequency as  $A$ and $B$ are complex matrices and
has substantial amplitude  for $\omega_e/k$
within the range of the speeds of the mean flow.
The response
of the time-dependent system is  highly amplified at the frequency 
of  the largest non-normal response
of the time-independent operator, which is at $\omega_e/k \approx 0.48$; and in sideband frequencies
that are separated from the frequency of the main maximum peak  by integral multiples of $\omega_f$. 
The equivalent-normal response  (dashed blue line in Figure \ref{fig:Fig3}(b))
peaks at $\omega_e/k \approx1$, which corresponds to the frequency of
the least damped mode of $A$,
while the variance of the time-dependent dynamics peaks at $\omega_e/k = 0.48$, which 
corresponds to the frequency of maximum non-normality of the time-independent operator,  
indicating that it is the non-normal structures of the time-independent 
dynamics that primarily influences both the time-independent and the time-dependent response.    Also note that,
 the response peaks 
irrespectively of the frequency of the time modulation, $\omega_f$,
at frequency $\omega_e/k \approx 0.48$ and not at twice the frequency, as was the case in the effectively normal
Mathieu oscillator.  The largest Lyapunov exponent of the flow
also determines the width of the Lorentzian peaks of 
the time-dependent response, as  shown in Figure \ref{fig:Fig3}(b). 
%\vskip-1.0in
\section{Conclusions}
This report describes  a study of the interaction among linear non-normal 
growth, fluctuation-fluctuation nonlinearity, and time dependence of the streamwise mean flow  in the dynamics of turbulence in shear flow.  %  In the first part of the study 
In the first part of this study an experiment is described in which the nonlinear fluctuation-fluctuation term in a streamwise mean and 
fluctuation from the streamwise mean partition of the Navier-Stokes equations 
are modified to allow variation in the magnitude of the fluctuation-fluctuation nonlinearity by a factor $\alpha$.  
Three turbulent regimes were found as $\alpha$ was varied: a parametric regime  that includes 
RNL turbulence
at $\alpha=0$ and
Navier-Stokes turbulence
at $\alpha=1$;
transition to a feedback regenerative regime with a constant streamwise mean flow  at $\alpha \approx 10$, and finally a laminar regime for  $\alpha>20$.  
Having identified the parametric regime as supporting the turbulent state in 
Navier-Stokes and RNL turbulence, we turned in the second part of this study to an examination of the mechanism of parametric growth using  time-dependent resolvent analysis.   
%The results indicate that  the eigenmodes of the 
%mean operator constitute the underlying structure inherited 
%by the structurally varying covariant Lyapunov vectors which in time-dependent flows
%assume a role analogous to that of the eigenvectors in time-independent 
%operator stability analysis.  Moreover, 
The non-normality of the mean 
operator and, in particular,  the frequency 
of maximum optimal response of the time-independent operator
were found to determine the central and sideband frequencies of the time-dependent system and  the structure of the optimal response to 
harmonic excitation. Also, the non-normal growth mechanism was found using Hill matrix analysis to be partitioned in a system-dependent manner between contributions from the non-normality of the mean and that of the time-dependent components of the operator (as in Fig \ref{fig:Fig2}).

Unlike the cases treated in this report,
time variations of the mean flow in turbulence are 
broadband in frequency. From our results,  we expect that
time dependence will amplify primarily the highly non-normal structures of the time-independent flow,
so that while with broadband mean flow time dependence the response will be
smoother than the response that arises  when the fluctuations of the mean flow 
are confined to distinct frequencies, it will  still be characterized primarily by the structure of the most amplified structure
of the time-independent dynamics.  
%One implication of this spreading of the influence 
%of the most non-normally amplified structure over frequencies is that while time dependence
%amplifies it also reduces the variability of the primary response structures across frequencies,
%arguing that spectral proper orthogonal decomposition (SPOD) methods should produce robust structures.
%

%===================================================================
\subsection*{Acknowledgments}
%The authors acknowledge use of computational resources from the Certainty cluster
%awarded by the National Science Foundation to CTR. 
%The authors acknowledge use of computational resources from the Yellowstone cluster awarded by the National Science Foundation to CTR.
We  thank Prof. Georgios Rigas, Prof.  Adrian Lozano-Dur\'an, Prof. Javier Jim\'enez and Dr. Mario di Renzo for their useful comments and discussions. 
%Marios-Andreas Nikolaidis gratefully acknowledges the support of the Hellenic Foundation for Research and
%Innovation (HFRI) and the General Secretariat for Research and Technology (GSRT).
%Brian F. Farrell was partially supported by NSF AGS-1640989.George Rigas, Adrian Lozano-Dur{\'a}n Javier Jimenez Mario di Renzo Insert acknowledgments here. 

%===================================================================
% THE BIBLIOGRAPHY
%=====================================================================
\bibliographystyle{ctr}
%\bibliography{brief}
%\bibliography{../../bibfile/basic_references_20110713}
%\bibliography{../../../bibfile/basic_references}
%\end{document}

\end{document}